\documentclass[]{amsart}
\usepackage[foot]{amsaddr}
\usepackage[a4paper,top=2.5cm,bottom=3cm,inner=3cm,outer=3cm]{geometry}

\usepackage[utf8]{inputenc}

\usepackage{tikz}
\usetikzlibrary{decorations.pathreplacing, decorations.markings, arrows.meta}
\usetikzlibrary{arrows}

\usepackage{amsmath,amsthm}
\usepackage{amsfonts,amssymb,mathrsfs}

\usepackage[bookmarks=false,breaklinks=true]{hyperref}
\usepackage{float}
\usepackage{graphicx}
\usepackage{color}

\usepackage{epigraph}
\definecolor{myurlcolor}{rgb}{0,0,0.7}
\usepackage{hyperref}
\hypersetup{colorlinks,
linkcolor=myurlcolor,
citecolor=myurlcolor,
urlcolor=myurlcolor}
\usepackage[mathscr]{euscript}

\definecolor{darkyellow}{rgb}{0.98,0.98,0}
\definecolor{darkgreen}{rgb}{0,0.9,0}

\definecolor{darkred}{rgb}{0.5,0,0} 

\newcommand{\R}{{\mathbb R}}  

\newcommand{\define}[1]{{\bf \boldmath{#1}}}

\theoremstyle{definition}

        \newcommand{\be}{\begin{equation}}
        \newcommand{\ee}{\end{equation}}
        \newcommand{\ba}{\begin{eqnarray}}
        \newcommand{\ea}{\end{eqnarray}}
        \newcommand{\ban}{\begin{eqnarray*}}
        \newcommand{\ean}{\end{eqnarray*}}
        \newcommand{\barr}{\begin{array}}
        \newcommand{\earr}{\end{array}}

\begin{document}
\title{The Inverse Cube Force Law}
\author[Baez]{John C.\ Baez} 
\address{School of Mathematics, University of Edinburgh, James Clerk Maxwell Building, Peter Guthrie Tait Road, Edinburgh, UK EH9 3FD}
\maketitle

\begin{figure}[h]
\includegraphics[width = 14em]{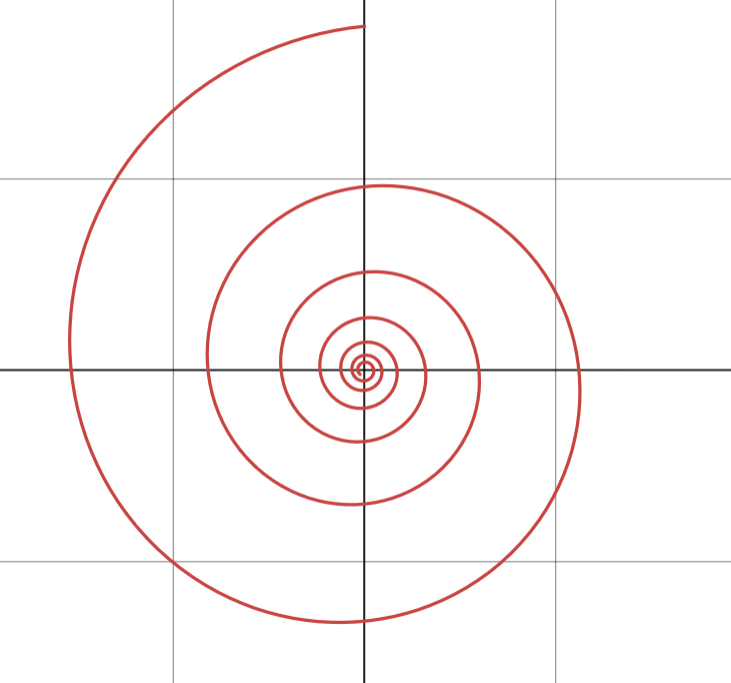} 
\caption{A particle spiraling into the origin in an inverse cube force.}
\label{figure_1}
\end{figure}

Newton's \textsl{Principia} is famous for its investigations of the inverse square force law for gravity.  But in this book Newton also did something that remained little-known until fairly recently \cite{Chandrasekhar,Wikipedia}.  He figured out what kind of central force exerted upon a particle can rescale its angular velocity by a constant factor without affecting its radial motion.  This turns out to be a force obeying an inverse \emph{cube} law.

Given a particle in Euclidean space, a \define{central force} is a force that points toward or away from the origin and depends only on the particle's distance from the origin.    If the particle's position at time \(t\) is \(\mathbf{r}(t) \in \R^n\) and its mass is some number \(m > 0\), we have
\[      m \, \ddot{\mathbf{r}}(t) = F(r(t))  \,\hat{\mathbf{r}}(t), \]
where \(\hat{\mathbf{r}}(t)\) is a unit vector pointing outward from the origin at the point \(\mathbf{r}(t)\).  A particle obeying this equation always moves in a plane through the origin, so we can use polar coordinates and write the particle's position as $\bigl(r(t), \theta(t)\bigr)\).   With some calculation one can show the particle's distance from the origin, $r(t)$, obeys
\begin{equation}
\label{eq:1}
m \ddot r(t) = F(r(t)) + L^2/mr(t)^3 . 
\end{equation}
Here \(L = mr(t)^2 \dot \theta(t)\), the particle's \define{angular momentum}, is constant in time.  The second term in Equation \eqref{eq:1} says that the particle's distance from the origin changes as if there were an additional force pushing it outward.  This is a ``fictitious force'', an artifact of working in polar coordinates.   It is called the \define{centrifugal force}.  And it obeys an inverse cube force law!    

This explains Newton's observation.   Let us see why.   Suppose that we have two particles moving in two different central forces \(F_1\) and \(F_2\), each obeying a version of Equation (\ref{eq:1}), with the same mass \(m\) and the same radial motion \(r(t)\), but different angular momenta \(L_1\) and \(L_2\).  Then we must have
\[       F_1(r(t)) + L_1^2/mr(t)^3  =  F_2(r(t)) + L_2^2/mr(t)^3 . \] 
If the particle's angular velocities are proportional then \(L_2 = kL_1\) for some constant \(k\), so  
\[        F_2(r_1(t)) - F_1(r(t)) = (k^2 - 1)L_1^2/mr(t)^3  .\]
This says that \(F_2\) equals \(F_1\) plus an additional inverse cube force.

A particle's motion in an inverse cube force has curious features.  First compare Newtonian gravity, which is an attractive inverse square force, say \(F(r) = -c/r^2\) with \(c > 0\).   In this case we have
\[
   m \ddot r(t) = -c/r(t)^2 + L^2/mr(t)^3 . 
\]
Because \(1/r^3\) grows faster than \(1/r^2\) as \(r \downarrow 0\), as long as the angular momentum \(L\) is nonzero the repulsion of the centrifugal force will beat the attraction of gravity for sufficiently small \(r\), and the particle will not fall in to the origin.   The same is true for any attractive force \(F(r) = -c/r^p\)  with \(p < 3\).  But an attractive inverse cube force can overcome the centrifugal force and make a particle fall in to the origin.    

In fact there are three qualitatively different possibilities for the motion of a particle in an attractive inverse cube force \(F(r) = -c/r^3\), depending on the value of $c$.  With work \cite{Grossman} we can solve for $1/r$ as a function of $\theta$ (which is easier than solving for $r$).  There are three cases depending on the value of
 \[    \omega^2 = 1 - cm/L^2 ,  \]
vaguely analogous to the elliptical, parabolic and hyperbolic orbits of a particle in an inverse square force law:
 \[  \frac{1}{r(\theta)}  = \left\{ \begin{array}{lcl}
 A \cos(\omega \theta) + B \sin(\omega \theta) & \text{if} & \omega^2 > 0 \\ [3pt]
 A + B \theta & \text{if} & \omega = 0 \\  [3pt]
 A e^{|\omega| \theta} + B e^{-|\omega| \theta}  & \text{if} & \omega^2 < 0. 
 \end{array} \right.
 \]
 The third case occurs when the attractive inverse cube force is strong enough to overcome the centrifugal force: \(c > L^2/m\).  Then the particle can \emph{spiral in to its doom}, hitting the origin in a finite amount of time after infinitely many orbits.   An example is shown in Figure \ref{figure_1}.  
 
All three curves are called \define{Cotes spirals}, after Roger Cotes' work on the inverse cube force law, published posthumously in 1722.   Cotes seems to have been the first to compute the derivative of the sine function.   After Cotes' death at the age of 33, Newton supposedly said ``If he had lived we would have known something'' \cite{Gowing}.

The subtlety of the inverse cube force law is greatly heightened when we study it using quantum rather than classical mechanics  \cite{GTV}.  Here if $c$ is too large the theory is ill-defined, because there is no reasonable choice of self-adjoint Hamiltonian.   If $c$ is smaller the theory is well-behaved.   But at a certain borderline point it exhibits a remarkable property: spontaneous breaking of scaling symmetry.  I hope to discuss this in my next column.


\begin{thebibliography}{33}

\bibitem{Chandrasekhar} S.\ Chandrasekhar, \textsl{Newton's Principia for the Common Reader}, Oxford U.\ Press, Oxford, 1995, pp.\ 183--200.



\bibitem{GTV} D.\ M.\ Gitman, I.\ V.\ Tyutin and B.\ L.\ Voronov, Self-adjoint extensions and spectral analysis in Calogero problem,  \textsl{J.\ Phys.\ A} \textbf{43} (14) (2010),  145205.  Also available at \href{http://arxiv.org/abs/0903.5277}{arXiv:0903.5277}.

\bibitem{Gowing} R.\ Gowing, \textsl{Roger Cotes---Natural Philosopher}, Cambridge U.\ Press, Cambridge, 2002.

\bibitem{Grossman} N.\ Grossman, \textsl{The Sheer Joy of Celestial Mechanics}, Birkh\"auser, Basel, 1996, p.\ 34.


\bibitem{Wikipedia} Newton's theorem of revolving orbits, Wikipedia.  Available at \href{https://en.wikipedia.org/wiki/Newton's_theorem_of_revolving_orbits}{https://en.wikipedia.org/wiki/Newton's$\underline{\;\;}$ theorem$\underline{\;\;}$of$\underline{\;\;}$revolving$\underline{\;\;}$orbits}.

\end{thebibliography}
\end{document}